\documentclass[article,english,superscriptaddress,showpacs,showkeys]{revtex4-2}
\usepackage[T1]{fontenc}
\usepackage[utf8]{inputenc}
\usepackage{color}
\usepackage{babel}
\usepackage{array}
\usepackage{units}
\usepackage{amsmath}
\usepackage{amssymb}
\usepackage{graphicx}
\usepackage{bm}
\usepackage{todonotes}
\usepackage[pdfusetitle,
bookmarks=true,bookmarksnumbered=false,bookmarksopen=true,bookmarksopenlevel=2,
breaklinks=true,pdfborder={0 0 0},pdfborderstyle={},backref=false,colorlinks=true]
{hyperref}
\hypersetup{
	citecolor=blue,linkcolor=blue,urlcolor=blue}



\begin{document}
	
	\title{Single-quantum annihilation of positrons at low energies}
	
	\author{Peter.A. Krachkov}\email{P.A.Krachkov@inp.nsk.su}
	\author{Simon.V. Sorokin}\email{S.Sorokin1@g.nsu.ru}
	\affiliation{Budker Institute of Nuclear Physics, SB RAS, Novosibirsk, 630090, Russia}
	\affiliation{Novosibirsk State University, 630090 Novosibirsk, Russia}
	
	\date{\today}
\begin{abstract}
	We discuss the single-quantum positron annihilation with a bound electron in the near threshold region. The angular distribution and the total cross section are considered. We obtain a simple analytical expression for the Coulomb potential and for the screened potential. It is worth noting that the obtained results for the screened potential are universal and independent of the explicit form of the atomic potential. It is shown that screening significantly increases the cross section. We also obtain the analogous results for bound-free $e^+e^-$ photoproduction.
\end{abstract}
	
\maketitle

\section{Introduction}\label{sec1}

Single-quantum annihilation (SQA), a fundamental process in quantum electrodynamics, was first investigated by Fermi and Uhlenbeck \cite{Fermi1933}. Using non-relativistic wave functions, they estimated total cross sections for the $K$- and $L$-shell electrons. Later, several works \cite{Bhabha1934, Bethe1935} presented analytical results for the $K$-shell in the Born approximation. Subsequent works \cite{Jaeger1936, Johnson1964, Johnson1967} were devoted to the numerical computation of the differential and total cross sections using exact solutions of the Dirac equation in the Coulomb field. However, the expressions obtained in these works have complicated structure. Significantly more compact analytical expressions for cross sections were obtained \cite{Moroi1970} using modified Furry-Sommerfeld-Maue wave functions \cite{Furry1934, Sommerfeld1935, Johnson1960}.

Experimental studies of this process have been underway since 1950 in the works of Merik \cite{Meric1950_1, Meric1950_2}. However, at that time, experimental accuracy did not allow to state with certainty the presence of the SQA channel. Only in 1961 a group of researchers from the MIT announced \cite{Sodickson1961} its reliable detection and measured the total cross sections in the field of heavy nuclei (atomic number $Z>70$). The following experiments were set in the University of Göttingen \cite{Langhoff1963, Weigmann1963, Friedrich1971} and Kyoto University \cite{Mazaki1968, Mukoyama1979} in the 1960s and 1970s. They were mainly conducted for nuclei with $Z>70$ or for relativistic positrons with a kinetic energy greater than 300 keV. The only exception is the work \cite{Friedrich1971}, where the SQA of low-energy positrons (kinetic energy less than 200 keV) with the electrons of iodine atoms ($Z=53$) was studied. However, the accuracy of their results does not allow to determine which of the theoretical models best correlates with the experiment. The most up-to-date research was conducted at Brookhaven National Laboratory in the 1990s \cite{Palathingal1991, Palathingal1995}. The accuracy of their measurements has increased significantly compared to previous experiments, but they also studied only relativistic positrons. 

Thus, the single-quantum annihilation of low-energy positrons in the field of light nuclei is of interest both from a theoretical and an experimental point of view, since it remains largely unexplored. In this paper, we present closed-form analytical expressions for cross sections for the $K$- and $L$-shell electrons. Here we assume that $Z\alpha \sim v_{q} \ll 1$ and $Z\alpha/v_{q} \sim 1$ ($\alpha$ denotes the fine-structure constant, $v_{q}$ represents positron velocity, and we use natural units where $\hbar=c=1$). Note that although we are considering a non-relativistic process, we should use relativistic wave functions for the electron and positron, since relativistic corrections make a significant contribution to the results. We use non-relativistic decomposition of exact solutions of the Dirac equation in the atomic field -- the method used in recent articles on the near-threshold pair photoproduction \cite{Krachkov2022, Krachkov2023}. 

The influence of atomic screening on the process of positron SQA at low energies remains almost unexplored \cite{Broda1972, Tseng1973}. In the present paper we obtain a simple analytical expression for the total cross section in screened potential. It is worth noting that this result is universal and independent of the explicit form of the atomic potential.

Finally, we obtain cross section for an associated process – near-threshold bound-free pair production.

\section{General discussion}\label{sec2}
The differential cross section for SQA for a single electron with principal quantum number $n$, orbital angular momentum $l$, total angular momentum $j$ and its projection $\mu \equiv j_{z}$, averaged over the polarization of the incoming positron and summed over the polarization of the final photon, is given by
\begin{align}\label{f1}
	d\sigma_{njl}=\frac{m\alpha}{2\pi v_{q}} \, \sum_{\lambda,\tau=\pm 1}^{} |\mathcal{M_{\lambda\tau}}|^{2} d\Omega\,,
\end{align}
where $m$ is the electron mass, $d\Omega$ is the solid angle corresponding to the direction of the photon momentum $\bm k$, and the matrix element $\mathcal{M_{\lambda\tau}}$ is
\begin{align}\label{MatrElem}
	\mathcal{M_{\lambda\tau}}=\int_{}^{}d^{3}r \text{ } e^{-i(\bm k \cdot \bm r)} \overline{V}^{(out)}_{\bm q,\,\tau}(\bm r)(\bm \gamma\cdot \bm e_{\lambda})U_{njl\mu}(\bm r)\,,
\end{align}
where  $\bm q$ and $\tau$ are positron momentum and spin projection, $\bm \gamma$ are the Dirac matrices, $\bm e_{\lambda}$ is the polarization vector for a photon, and $\lambda$ is its helicity. The Dirac wave function $U_{njl\mu}(\bm r)$ corresponds to a bound electron with quantum numbers $n$, $l$, $j$ and $\mu \equiv j_{z}$. The relativistic wave function for an incoming positron $V^{(out)}_{\bm q,\,\tau}(\bm r)$  at large distances contains a plane wave and a converging spherical wave.

At first, we will neglect the effect of the atomic screening. The effect of atomic screening on the process under consideration will be discussed below. This allows us to consider the wave functions of the incoming positron and bound electron as well-known relativistic Coulomb functions. The wave function of a bound electron has the form \cite{Landau4}
\begin{align}\label{U_el}
	U_{njl\mu}(\bm r)=\binom{f_{njl}(r) \text{ } \Omega_{jl\mu}(\bm n)}{-\sigma g_{njl}(r) \text{ } \Omega_{jl'\mu}(\bm n)}\,,
\end{align}
where
\begin{align}
	l=j+\frac{\sigma}{2}\,, ~~~~ l^{'}=j-\frac{\sigma}{2}\,, ~~~~ \sigma=\pm 1\,, \nonumber
\end{align}
$f_{njl}(r)$, $g_{njl}(r)$ are the radial functions, $\bm n=\bm r/r$, and $\Omega_{jl\mu}(\bm n)$ are spherical spinors. The  discrete spectrum radial functions can be represented as
\begin{align}\label{El_rad_func}
	f_{njl}(r)=\frac{(2\Lambda)^{\frac{3}{2}}}{\Gamma(2\gamma+1)} &\left[ \frac{(m+ \varepsilon)\Gamma(2\gamma+n_{r}+1)} {4m\,\frac{m\eta}{\Lambda}(\frac{m\eta}{\Lambda}-\kappa)\,n_{r}!} \right]^{\frac{1}{2}} (2\Lambda r)^{\gamma-1}e^{-\Lambda r}      \nonumber\\
	&\times  \Big\{ \left( \frac{m\eta}{\Lambda}-\kappa \right)\,_1F_1(-n_{r};\,2\gamma+1;\,2\Lambda r) - n_{r\,1}F_1(1-n_{r};\,2\gamma+1;\,2\Lambda r)\Big\}\,,    \nonumber\\
	g_{njl}(r)=\frac{-(2\Lambda)^{\frac{3}{2}}}{\Gamma(2\gamma+1)} &\left[ \frac{(m- \varepsilon)\Gamma(2\gamma+n_{r}+1)}{4m\,\frac{m\eta}{\Lambda}(\frac{m\eta}{\Lambda}-\kappa)\,n_{r}!} \right]^{\frac{1}{2}}(2\Lambda r)^{\gamma-1}e^{-\Lambda r}          \\
	&\times  \Big\{ \left( \frac{m\eta}{\Lambda}-\kappa \right)\, _1F_1(-n_{r};\,2\gamma+1;\,2\Lambda r)+ n_{r \,1}F_1(1-n_{r};\,2\gamma+1;\,2\Lambda r)\Big\}\,,    \nonumber\\
	\eta = Z\alpha\,, \quad \kappa=\sigma \bigg( j+\frac{1}{2} &\bigg)\,, \quad \gamma=\sqrt{\kappa^{2}-\eta^{2}}\,, \quad \Lambda = \sqrt{m^{2}- \varepsilon^{2}}\,, \quad \varepsilon=m\left[ 1+\frac{\eta^{2}}{(\sqrt{\kappa^{2}-\eta^{2}}+n_{r})^{2}} \right]^{-\frac{1}{2}}\,. \nonumber
\end{align}
Here $\Gamma(z)$ is the Euler gamma function, $\text{}_1F_1(a;\, b;\, z)$ is the confluent hypergeometric function (Kummer function), $n_{r}$ is the radial quantum number of the electron, and $\varepsilon$ is its total energy. It is worth noting that the radial quantum number can take values $n_{r}=0,1,2,...$ at $\sigma=-1$ and $n_{r}=1,2,...$ at $\sigma=+1$. However, in both cases, the principal quantum number can be expressed in terms of the total angular momentum and the radial quantum number $$n=n_{r}+j+\frac{1}{2}\,.$$
The wave function of a incoming positron cannot be represented in a closed form. It has the form of an infinite sum over partial waves with a certain values of the orbital angular momentum $L$, the total angular momentum $J$ and its projection $M\equiv J_{z}$. The form of this sum is determined by the requirement that the wave function at large distances contains a plane wave and a converging spherical wave. Based on this, the positron wave function can be represented as
\begin{align}\label{V_pos}
	V^{(out)}_{\bm q,\,\tau}(\bm r)=\frac{4\pi}{2q}\sum_{L,\, M}^{}\frac{i^{-L}Y^{*}_{L,\, M-\tau/2}(\bm n_{\bm q})}{\sqrt{2L+1}} &\Bigg[ e^{-i\Delta^{(+)}_{L}}\sqrt{L+\frac{1}{2}+\tau M} \, \binom{G^{(+)}_{L} \text{ } \Omega_{L+1/2,\, L+1, \, M}}{-F^{(+)}_{L} \text{ } \Omega_{L+1/2,\, L, \, M}}    \\ 
	&+\tau e^{-i\Delta^{(-)}_{L}}\sqrt{L+\frac{1}{2}-\tau M}\, \binom{G^{(-)}_{L} \text{ } \Omega_{L-1/2,\, L-1, \, M}}{F^{(-)}_{L} \text{ }  \Omega_{L-1/2,\, L, \, M}} \Bigg]\,, \nonumber
\end{align}
where $F^{(\pm)}_{L}(r)$, $G^{(\pm)}_{L}(r)$ are the radial functions, $Y_{LM}(\bm n_{\bm q})$ are spherical harmonics, $\bm n_{q} = \bm q/q$ and $\Delta^{(\pm)}_{L}$ are the phases determined by the behavior of the radial functions at large distances
\begin{align}
	&
	\begin{pmatrix}
		F^{(\pm)}_{L}(r)\\
		G^{(\pm)}_{L}(r)
	\end{pmatrix} \underset{r\to\infty}{\to} \frac{2}{r\sqrt{2\varepsilon_{q}}}
	\begin{pmatrix}
		\sqrt{\varepsilon_{q}+m}\,\sin(qr-\frac{\pi L}{2}+\Delta^{(\pm)}_{L})\\
		-\sqrt{\varepsilon_{q}-m}\,\cos(qr-\frac{\pi L}{2}+\Delta^{(\pm)}_{L})
	\end{pmatrix}\,.  \nonumber
\end{align}
Indices $(\pm)$ indicate the relationship between $J$ and $L$: $J=L\pm 1/2$. Radial functions $F(r)$ and $G(r)$ have the form
\begin{align}\label{Pos_rad_func}
	F^{(\pm)}_{L}(r)=\frac{\sqrt{2}}{r} \sqrt{\frac{\varepsilon_{q}+ m}{\varepsilon_{q}}}e^{-\frac{\pi\nu_{q}}{2}}\frac{|\Gamma(\gamma+1-i\nu_{q})|}{\Gamma(2\gamma+1)}(2q r)^{\gamma} \, &\text{Re}\!\left\{ e^{i(qr+\xi)}\,_1F_1(\gamma+i\nu_{q};\,2\gamma+1;\,-2iq r)\right\}\,,    \nonumber\\
	G^{(\pm)}_{L}(r)=\frac{\sqrt{2}}{r} \sqrt{\frac{\varepsilon_{q}- m}{\varepsilon_{q}}}e^{-\frac{\pi\nu_{q}}{2}}\frac{|\Gamma(\gamma+1-i\nu_{q})|}{\Gamma(2\gamma+1)}(2q r)^{\gamma} \, &\text{Im}\!\left\{ e^{i(qr+\xi)}\,_1F_1(\gamma+i\nu_{q};\,2\gamma+1;\,-2iq r)\right\}\,,   \\
	\nu_{q}=\frac{\eta\varepsilon_{q}}{q}\,, \quad e^{2i\xi}=\frac{\kappa^{(\pm)}+i\nu_{q}\frac{m}{\varepsilon_{q}}}{\gamma+i\nu_{q}}\,,&  \quad \kappa^{(\pm)}=\mp\left( J+\frac{1}{2} \right)\,,         \nonumber
\end{align}
where the definition of $\gamma$ coincides with that was defined in (\ref{El_rad_func}) up to the replacement of $j\to J$.

In order to take into account the effect of atomic screening on the process under consideration, we note that in the near threshold region, the matrix element \eqref{MatrElem} has the following property. The integral converges at small distances 
$$r\sim \frac{1}{k}\sim \lambda_{c}\,,$$
where $\lambda_{c}$ is the Compton wavelength for the electron. The effect of screening is significant only at distances of the screening radius  $r\sim r_{scr}\sim a_{\text{B}}Z^{-1/3}\gg \lambda_{c}$, where $a_{\text{B}}=\lambda_{c}/\alpha$ is the Bohr radius. Therefore, near the origin, the screened wave functions differ from the Coulomb ones only by the normalization factor. Thus, Coulomb wave functions with a modified normalization factor are an appropriate approximation of them. For convenience, we will separate the amplitude and phase of this factor at a certain value of the orbital angular momentum

\begin{align}\label{Scr_wave_func}
	&V^{scr}_{L}(\bm r)=\sqrt{\widetilde{A}_{L}(q)} \text{ } V_{L} (\bm r)\, e^{i\chi_{L}}\,, \quad U^{scr}_{nlj}(\bm r)=\sqrt{B_{nl}} \text{ }U_{nlj}(\bm r)\, e^{i\phi_{nlj}}\,, \nonumber \\
	&\widetilde{A}_{L}(q)=\lim_{r \to 0} \left| \frac{F^{scr}_{L}(r)}{F_{L}(r)} \right|^{2}\,, \quad \quad \quad ~~\, B_{nl}=\lim_{r \to 0} \left| \frac{f^{scr}_{nlj}(r)}{f_{nlj}(r)} \right|^{2}\,,
\end{align}
where the radial functions $F_{L}(r)$ and $f_{nlj}(r)$ are defined in (\ref{Pos_rad_func}) and (\ref{El_rad_func}), respectively. Note that with the required accuracy the normalization factors $A$ and $B$ do not depend  on total angular momentum. We will also redefine the factor for the positron wave function by introducing a factor
\begin{align}\label{A_redifine}
	\widetilde{A}_{L}(q)=\mathcal{F}_{L}A_{L}(q)\,, 
\end{align}
where $\mathcal{F}_{L}$ is the Sommerfeld-Gamov-Sakharov factor \cite{Sommerfeld2, Gamow1928, Sakharov1948} for the positron partial wave with orbital angular momentum $L=0$. We will define it later during the calculation of matrix elements.

It is important to note that this approach does not depend on the screening model, since only its most general properties were used. But in order to get quantitative results, it is necessary to use a specific representation of the potential. As an example, we use in this work the Thomas-Fermi approximation. Despite the fact it is the simplest screening model, it still has no analytical expression for the potential. Therefore, Molière constructed an approximation of it using experimental data \cite{Moliere1947}
\begin{align}\label{Moliere}
	V_{M}(r)=\frac{Z\alpha}{r}\left[ 0.1\text{ }e^{-6 \, r/a}+0.55\text{ }e^{-1.2 \, r/a}+0.35\text{ }e^{-0.3 \, r/a_{}} \right]\,,
\end{align}
where
$$a=\frac{1}{2}\left( \frac{3\pi}{4} \right)^{2/3}a_{\text{B}}Z^{-1/3}\sim \frac{121}{137}a_{\text{B}}Z^{-1/3}$$
is the Thomas-Fermi screening radius.
 
The influence of static potential, polarization potential and exchange interaction on the wave function at small distance was studied in detail in \cite{Krachkov2023}. We will not touch upon these issues here.

\section{Calculation of matrix elements}\label{sec3}
\subsection{j = 1/2, l = 0}\label{sec31}
First, we consider the case of SQA in Coulomb field with the electron in the $nS_{1/2}$ state. Note that in this section we assume $\Omega_{jlm}=\Omega_{jlm} (\bm n)$. To obtain the cross section in the leading order according to parameter $\eta=Z\alpha$, it is necessary to take into account three components of the positron wave function. The first one corresponds to the positron partial wave with $J=1/2$, $L=0$, $M_{1}=\pm 1/2$
\begin{align}\label{V_1}
	V_{1}=\frac{4\pi}{2q} \, \sum_{M_{1}}^{} \frac{1}{\sqrt{4\pi}}e^{-i\Delta^{(+)}_{0}} \sqrt{1/2+\tau M_{1}} \, \binom{G^{(+)}_{0}(r) \text{ } \Omega_{1/2,\, 1, \, M_{1}}}{-F^{(+)}_{0}(r) \text{ } \Omega_{1/2,\, 0, \, M_{1}}}\,.
\end{align}
The second one is the partial wave with $J=1/2$, $L=1$, $M_{2}=\pm 1/2$
\begin{align}\label{V_2}
	V_{2}=-\frac{4\pi i \tau}{2q} \, \sum_{M_{2}}^{} Y^{*}_{1, \, M_{2}-\tau/2}(\bm n_{\bm q})e^{-i\Delta^{(-)}_{1}} \sqrt{\frac{3/2-\tau M_{2}}{3}} \, \binom{G^{(-)}_{1}(r) \text{ } \Omega_{1/2,\, 0, \, M_{2}}}{F^{(-)}_{1}(r) \text{ } \Omega_{1/2,\, 1, \, M_{2}}}\,.
\end{align}
And the third one is the partial wave with $J=3/2$, $L=1$, and $M_{3}=\pm 1/2, \pm 3/2$
\begin{align}\label{V_3}
	V_{3}=-\frac{4\pi i}{2q} \, \sum_{M_{3}}^{} Y^{*}_{1,\, M_{3}-\tau/2}(\bm n_{\bm q})e^{-i\Delta^{(+)}_{1}} \sqrt{\frac{3/2+\tau M_{3}}{3}} \, \binom{G^{(+)}_{1}(r) \text{ } \Omega_{3/2,\, 2, \, M_{3}}}{-F^{(+)}_{1}(r) \text{ } \Omega_{3/2,\, 1, \, M_{3}}}\,.
\end{align}
It is worth noting that the contribution to the cross section from $M_{3}=\pm 3/2$ will be zero. The difference between phases $\Delta^{(\pm)}_{1}$ is negligibly small $\Delta^{(+)}_{1}-\Delta^{(-)}_{1}= \mathcal{O}(\eta^{2})$.
Therefore, these phases can be considered in the non-relativistic approximation $\Delta^{(+)}_{1}=\Delta^{(-)}_{1}=\Delta_{1}$. The final expressions for the cross sections will not depend on the phases at all. 

The main contribution to the matrix element is determined by the distances $r \sim \lambda_{c}$. Therefore, we expand the wave functions in $\Lambda r, q r \ll 1$ up to second order. After some tedious calculations, we obtain squared matrix elements for $\lambda,\tau=\pm 1$
\begin{align}\label{M_ns1/2}
&|\mathcal{M}_{\pm\pm}|^2= \delta_{\mu,\,\mp 1/2} \frac{9\pi^4}{8} \frac{\eta^7}{m^3 n^3} \mathcal{F}_{0} \left| Y_{0,\, 0} \right|^2\,, \nonumber \\
&|\mathcal{M}_{\pm\mp}|^2=\delta_{\mu,\,\mp 1/2}\frac{4\pi^2}{3 \nu_{q}^{2}} \frac{\eta^7}{m^3 n^3} \mathcal{F}_{1} \left| Y_{1,\, \pm 1}(\bm n_{\bm q}) \right|^2\,.   
\end{align}
Here $\nu_{q}=\eta/v_{q}$ coincides with that was defined in (\ref{Pos_rad_func}),  $\mathcal{F}_{0}$ and $\mathcal{F}_{1}$ are the Sommerfeld-Gamov-Sakharov factors for the positron $s$-wave and $p$-wave
\begin{align}\label{SGS_Fact}
	\mathcal{F}_{0}=\frac{2\pi\nu_{q}}{e^{2\pi\nu_{q}}-1}\,,\quad \mathcal{F}_{1}=\mathcal{F}_{0}(1+\nu^{2}_{q})\,.
\end{align}
Due to the non-relativistic approximation of phases, there is no interference between the partial waves of the positron. Thus, in the leading order, $|\mathcal{M}_{\pm\pm}|^2\,$ corresponds to partial wave with $L=0$, and $|\mathcal{M}_{\pm\mp}|^2$ corresponds to $L=1$.

As mentioned above, the screened wave functions differ from Coulomb ones only by a normalization factor (\ref{Scr_wave_func}). Therefore, to take into account the screening of the electron wave function in the $nS_{1/2}$ state, it is necessary to multiply squared matrix elements by the coefficient $B_{nS}$. Taking into account the screening of the positron wave function will lead to the replacement of Sommerfeld-Gamov-Sakharov factors $\mathcal{F}_{0}$ and $\mathcal{F}_{1}=\mathcal{F}_{0}(1+\nu^{2}_{q})$ by coefficients $A_{0}(q)$ and $A_{1}(q)$. As a result, we have
\begin{align}\label{ScrM_ns1/2}
	&|\mathcal{M}_{\pm\pm}^{scr}|^2= \delta_{\mu,\,\mp 1/2} \frac{9\pi^4}{8} \frac{\eta^7}{m^3 n^3}  A_{0}(q) B_{nS} \left| Y_{0,\, 0} \right|^2 \,, \nonumber \\
	&|\mathcal{M}_{\pm\mp}^{scr}|^2= \delta_{\mu,\,\mp 1/2} \frac{4\pi^2}{3 \nu_{q}^{2}} \frac{\eta^7}{m^3 n^3} A_{1}(q) B_{nS} \left| Y_{1,\, \pm1}(\bm n_{\bm q}) \right|^2\,.   
\end{align}

\subsection{j = 1/2, l = 1}\label{sec32}
For an electron in the $nP_{1/2}$ state to obtain the leading order cross section it is necessary to take into account only one component of the positron wave function -- $V_{1}$, defined in (\ref{V_1}). As a result of similar calculations, we obtain
\begin{align}\label{M_np1/2}
	&|\mathcal{M}_{\pm\pm}|^2 = \delta_{\mu,\,\mp 1/2} \frac{2\pi}{9} \frac{\eta^7}{m^3} \frac{n^{2}-1}{n^{5}} \mathcal{F}_{0}\,,  \\
	&|\mathcal{M}_{+-}|^2 =|\mathcal{M}_{-+}|^2 =0\,.    \nonumber
\end{align}
It can be seen that the expression does not depend on the vector $\bm n_{\bm q}$, which leads to an isotropic angular distribution. Also note that the phases are again not included in the expression for cross sections.

In the screened potential, the matrix elements are transformed similarly to what was shown in \ref{sec31}. Then for non-zero $|\mathcal{M}_{\lambda\tau}^{scr}|^2$ we have
\begin{align}\label{ScrM_np1/2}
	&|\mathcal{M}_{\pm\pm}^{scr}|^2= \delta_{\mu,\,\mp 1/2} \frac{2\pi}{9} \frac{\eta^7}{m^3} \frac{n^{2}-1}{n^{5}} A_{0}(q) B_{nP}\,.
\end{align}

\subsection{j = 3/2, l = 0}\label{sec33}
As in the previous section, an electron in the $nP_{3/2}$ state it is necessary to take into account only $V_{1}$, defined in (\ref{V_1}). This immediately indicates that the angular distribution will be isotropic, and the phases will not be included in the expression for the cross section. Note that the non-zero contribution to the matrix element arises only from an electron with  $\mu=j_z=\pm \frac{1}{2}$. As a result of similar calculations, we obtain
\begin{align}\label{M_np3/2}
	&|\mathcal{M}_{\pm\pm}|^2= \delta_{\mu,\,\mp 1/2} \frac{\pi}{9} \frac{\eta^7}{m^3}  \frac{n^{2}-1}{n^{5}} \mathcal{F}_{0}\,, \\
	&|\mathcal{M}_{+-}|^2=|\mathcal{M}_{-+}|^2=0\,.    \nonumber
\end{align}
Applying the already familiar substitutions, we get non-zero squared matrix elements in the screened potential
\begin{align}\label{ScrM_np3/2}
	&|\mathcal{M}_{\pm\pm}^{scr}|^2= \delta_{\mu,\,\mp 1/2} \frac{\pi}{9} \frac{\eta^7}{m^3} \frac{n^{2}-1}{n^{5}} A_{0}(q) B_{nP}\,.
\end{align}

\section{Discussion of the results}\label{sec4}
\subsection{Coulomb potential}\label{sec41}
Using Eqs. (\ref{f1}) and (\ref{M_ns1/2}), we will write down the differential cross section for a single electron in the $nS_{1/2}$ state
\begin{align}\label{DiffCS_ns1/2}
	d\sigma_{nS_{1/2}}=\frac{1}{4}\frac{1}{n^{3}}\frac{\alpha(Z\alpha)^{5}}{m^{2}v_{q}}  \frac{2\pi\nu_{q}}{e^{2\pi\nu_{q}}-1} \bigg[\underbrace{\frac{9\pi^{2}}{16}(Z\alpha)^{2}}_{L=0}+ (1+\nu^{2}_{q})\underbrace{\left[\bm v_{q} \times \bm n_{k}  \right]^{2}}_{L=1}\bigg]d\Omega\,,
\end{align}
where $\nu_{q}=(Z\alpha)/v_{q}$. Note that the dependence on the principal quantum number $n$ is factorized as $1/n^{3}$. Note also the Sommerfeld-Gamov-Sakharov factors $\mathcal{F}_{0}$ and $\mathcal{F}_{1}$ before the components with $L=0$ and $L=1$ (they are defined in (\ref{SGS_Fact})). They significantly suppress the cross section at low energies. 

After integration over the angles, we obtain the total cross section
\begin{align}\label{TotCS_ns1/2}
	\sigma_{nS_{1/2}}=\frac{2\pi}{3}\frac{1}{n^{3}}\frac{\alpha(Z\alpha)^{5}}{m^{2}v_{q}} \frac{2\pi\nu_{q}}{e^{2\pi\nu_{q}}-1}\left[ v^{2}_{q}+\left( \frac{27\pi^{2}}{32}+1 \right)(Z\alpha)^{2} \right]\,.
\end{align}

The differential cross sections for $p$-electrons in the leading order have a contribution only from the positron $s$-wave, therefore, the angular distribution is isotropic
\begin{align}\label{DiffCS_np}
	&d\sigma_{nP_{1/2}}=\frac{1}{9}\frac{n^{2}-1}{n^{5}}\frac{\alpha(Z\alpha)^{5}}{m^{2}v_{q}} \frac{2\pi\nu_{q}}{e^{2\pi\nu_{q}}-1} (Z\alpha)^{2}d\Omega\,, \nonumber \\ 
	&d\sigma_{nP_{3/2}}=(\delta_{\mu,1/2}+\delta_{\mu,-1/2}) \frac{1}{18} \frac{n^{2}-1}{n^{5}} \frac{\alpha(Z\alpha)^{5}}{m^{2}v_{q}} \frac{2\pi\nu_{q}}{e^{2\pi\nu_{q}}-1}(Z\alpha)^{2}d\Omega\,, 
\end{align}
where $\delta_{\mu,\,\pm 1/2}$ are the Kronecker symbols, reminding that only electrons with the projection $\mu=\pm 1/2$ participate in the annihilation process. The Sommerfeld-Gamov-Sakharov factor (\ref{SGS_Fact}) is also present in the formulae, and the dependence on the principal quantum number is again factorized. Total cross sections can be simply obtained by multiplying (\ref{DiffCS_np}) by $4\pi$
\begin{align}\label{TotCS_np}
	&\sigma_{nP_{1/2}}=\frac{4\pi}{9}\frac{n^{2}-1}{n^{5}}\frac{\alpha(Z\alpha)^{5}}{m^{2}v_{q}} \frac{2\pi\nu_{q}}{e^{2\pi\nu_{q}}-1} (Z\alpha)^{2}\,, \nonumber \\ 
	&\sigma_{nP_{3/2}}=(\delta_{\mu,\, 1/2}+\delta_{\mu,\, -1/2}) \frac{2\pi}{9}\frac{n^{2}-1}{n^{5}}\frac{\alpha(Z\alpha)^{5}}{m^{2}v_{q}} \frac{2\pi\nu_{q}}{e^{2\pi\nu_{q}}-1}(Z\alpha)^{2}\,.
\end{align}
Note that the order of all these cross sections is $\mathcal{O} \big(\alpha(Z\alpha)^{7}\big)$. As it turns out, in this approximation, the cross sections for electrons in any other state is $\operatorname{\scriptstyle\mathcal{O}}\big(\alpha(Z\alpha)^{7}\big)$, for this reason, they were not considered in this work.

Having cross sections for $s$- and $p$-electrons, it is not difficult to obtain total cross sections for the $K$- and $L$-shell electrons. There are two electrons in the $1S_{1/2}$ state on the $K$-shell, therefore, we obtain
\begin{align}\label{TotCS_K}
	\sigma_{K}=\frac{4\pi}{3}\frac{\alpha(Z\alpha)^{5}}{m^{2}v_{q}} \frac{2\pi\nu_{q}}{e^{2\pi\nu_{q}}-1}\left[ v^{2}_{q}+\left( \frac{27\pi^{2}}{32}+1 \right)(Z\alpha)^{2} \right]\,.
\end{align}
There are eight electrons on the $L$-shell: two in the $2S_{1/2}$ state, two in the $2P_{1/2}$ state, and four in the $2P_{3/2}$ state. But electrons with $\mu=\pm 3/2$ do not give a contribution to the leading order cross section, so only six electrons are taken into account in $\sigma_{L}$. Therefore, the total cross section for the $L$-shell electrons is
\begin{align}\label{TotCS_L}
	\sigma_{L}=\frac{\pi}{6}\frac{\alpha(Z\alpha)^{5}}{m^{2}v_{q}} \frac{2\pi\nu_{q}}{e^{2\pi\nu_{q}}-1}\left[ v^{2}_{q}+\left( \frac{27\pi^{2}}{32}+\frac{7}{4} \right)(Z\alpha)^{2} \right]\,.
\end{align}

In \cite{Johnson1964}, an exact numerical result is given for the cross section of SQA of positrons  with the $K$-shell electrons. Unfortunately, the numerical data from Ref.~\cite{Johnson1964} contain a small number of points in the low-energy region. For $\varepsilon_{q}=1.0625$ and $Z=47$ (silver), which corresponds to $v_q\sim0.35$, we obtain a $2\%$ difference between our and the numerical result. This discrepancy corresponds to the accuracy of the non-relativistic decomposition.

In Ref.~\cite{Krachkov2022} it was shown that the Furry-Sommerfeld-Maue approximation cannot give a right result for $e^+e^-$ pair production near threshold. We perform the  analogous analysis for SQA. We obtain that, using Furry-Sommerfeld-Maue approximation for both wave functions, one can obtain the right result for $d\sigma_{nP_{1/2}}$ and $d\sigma_{nP_{3/2}}$, see Eqs.~\eqref{DiffCS_np}. The result for the  $d\sigma_{nS_{1/2}}$ can not be obtained by the Furry-Sommerfeld-Maue approximation. Taking into account the above, our result is consistent with \cite{Moroi1970}, where results were obtained by the modified Furry-Sommerfeld-Maue approximation \cite{Johnson1960}. Our results are also consistent with the Born approximation, see \cite{Bethe1935}.

\subsection{Screened potential}\label{sec42}
We will investigate the effect of atomic screening on the total cross sections for the $K$- and $L$-shell electrons. Using the matrix elements (\ref{ScrM_ns1/2}), we obtain screened total cross section for the $K$-shell electrons
\begin{align}\label{Scr_TotCS_K}
	\sigma_{K}^{scr}=\frac{4\pi}{3}\frac{\alpha(Z\alpha)^{5}}{m^{2}v_{q}} B_{1S} \left[A_{0}(q) \frac{27\pi^{2}}{32}(Z\alpha)^{2} + A_{1}(q) \, v_{q}^{2} \right]\,.
\end{align}
Using matrix elements \eqref{ScrM_np1/2} and \eqref{ScrM_np3/2} one can obtain screened total cross sections for $p$-electrons. Combining the cross sections for the electrons in $2S_{1/2}$, $2P_{1/2}$ and $2P_{3/2}$ states the same way as it was done for Eq. (\ref{TotCS_L}), we obtain screened total cross section for the $L$-shell electrons
\begin{align}\label{Scr_TotCS_L}
	\sigma_{L}^{scr}=\frac{\pi}{6}\frac{\alpha(Z\alpha)^{5}}{m^{2}v_{q}}\left[A_{0}(q)\left(  \frac{27\pi^{2}}{32} B_{2S}+\frac{3}{4}B_{2P}\right)(Z\alpha)^{2} + A_{1}(q)B_{2S} \, v_{q}^{2} \right]\,.
\end{align}

\begin{figure}
	\centering
	\includegraphics[width=0.49\linewidth]{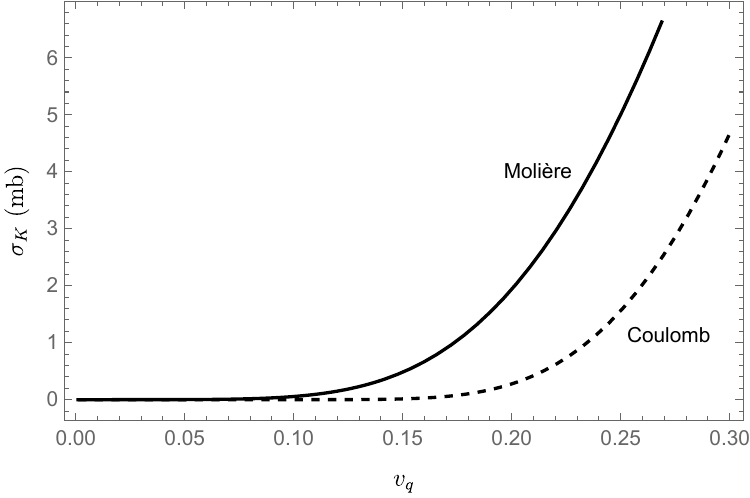}
	\includegraphics[width=0.49\linewidth]{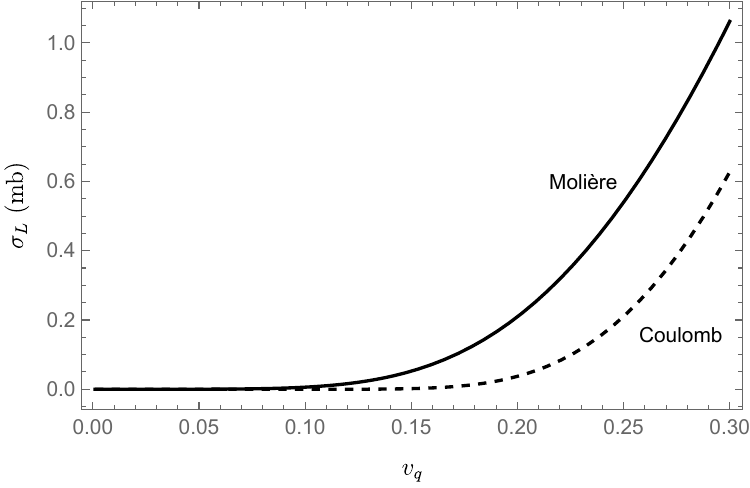}
	\caption{Total cross section of positron SQA  for the $K$-shell and $L$-shell electrons for silver ($Z=47$) in the Coulomb potential (dashed curve) and in the Molière potential (solid curve).}
	\label{fig:12}
\end{figure}

Fig. \ref{fig:12} shows SQA total cross sections for the $K$- and $L$-shell electrons for silver ($Z=47$) in the Coulomb potential and in the screened Molière potential (\ref{Moliere}). One sees that atomic screening increases the total cross section at any value of $v_{q}$.

\begin{figure}
	\centering
	\includegraphics[width=0.49\linewidth]{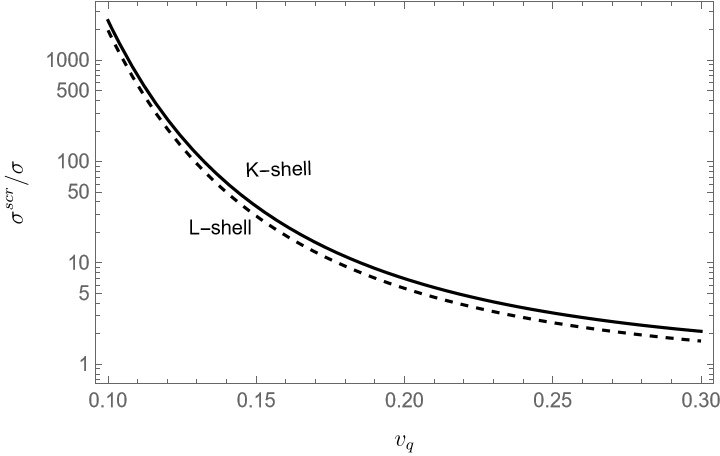}
	\caption{The ratio of the total cross section of SQA in the Molière potential to the total cross section in the Coulomb potential at $Z=47$. The ratio for the K-shell electrons (solid line) and for the L-shell electrons (dashed line) is given.}
	\label{fig:3}
\end{figure}

In order to estimate the effect of screening at $v_{q}<0.2$, in Fig. \ref{fig:3} we show the ratio of the screened cross section to the unscreened one  for  both $K$- and $L$-shell. One notices that at very low positron velocities, the cross section in the Coulomb field is significantly smaller than the screened cross section. This is associated with the fact that the Sommerfeld-Gamov-Sakharov factor tends to zero at low positron velocities, but the coefficients $A_{0}(q)$ and $A_{1}(q)$ tends to a non-zero constant.

\section{Bound-free pair production}\label{sec5}
Since the 1980s a significant number of papers have been devoted to the study of the related process -- bound-free pair production in the Coulomb field of the bare nuclei. However, both analytical \cite{Milstein1993, Belkacem1998, Bertulani2001, Sorensen2001, DiPiazza2012} and numerical \cite{Aste1994, Agger1997, Aste2008} results are presented at high-energies. The differential cross section for bound-free pair production, averaged over the polarization of the incoming photon and summed over the polarization of the final positron, is
\begin{align}\label{bfPP}
\sigma_{bf}=\frac{m\alpha v_{q}}{8\pi} \int_{}^{} d\Omega_{\bm q} \sum_{free}^{} \sum_{\lambda,\tau=\pm 1}^{} |\mathcal{M_{\lambda\tau}}|^{2} \,,
\end{align}
where squared matrix elements $|\mathcal{M}_{\lambda\tau}|^2$ are exactly the same which were calculated in Section \ref{sec3}, $d\Omega_{\bm q}$ is the solid angle corresponding to the direction of the positron momentum $\bm q$ and $\sum_{free}^{}$ means the summation over all free bound states at which the electron can be captured.

Since the cross sections of these two processes differ only by a simple kinematic factor, all the statements made above about SQA can be applied to photoproduction, also as a formulas for the different cross sections. We will not repeat all of them here. As an example, we will write the total cross section of bound-free $e^+e^-$ production in the field of the bare nuclei 
\begin{align}\label{bfPP_TotCS}
\sigma_{bf}=\frac{\pi\zeta(3)}{3}\frac{\alpha(Z\alpha)^{5}v_{q}}{m^{2}} \frac{2\pi\nu_{q}}{e^{2\pi\nu_{q}}-1}\left[v^{2}_{q}+\left( \frac{27\pi^{2}}{32}+2 - \frac{\zeta(5)}{\zeta(3)} \right)(Z\alpha)^{2}\right] \,,
\end{align}
or the $K$-shell total cross section
\begin{align}\label{bfPP_TotCS_K}
\sigma_{bf}^{K}=\frac{\pi}{3}\frac{\alpha(Z\alpha)^{5}v_{q}}{m^{2}} \frac{2\pi\nu_{q}}{e^{2\pi\nu_{q}}-1}\left[v^{2}_{q}+\left( \frac{27\pi^{2}}{32}+1\right)(Z\alpha)^{2}\right] \,,
\end{align}
where $\zeta(s)$ is the Riemann zeta function.

\section{Conclusion}\label{sec6}
In the present paper, we discuss the  single-quantum positron annihilation in the region $Z\alpha, v_q\ll1$ and $Z\alpha/v_q\sim 1$. We compute the differential and total cross section for $s$- and $p$-electrons in the leading order and show that the cross section with angular momentum of bound electron $l>1$ is suppressed. 

We also investigate the impact of screening on the differential and total cross sections and obtain  the simple analytical expression for Coulomb potential and for the screening potential. It is important to note that our results  are universal and independent of the explicit form of the atomic potential. We show that the screening effect can be taken into account by changing the value  of wave function normalization, while the phase of the wave function does not affect the result. To demonstrate the significance of the screening effect, we use the Molière approximation to the Thomas-Fermi potential and demonstrate that the effect of  screening  significantly increases the differential and total cross section. 

We also obtain the analogous results for bound-free $e^+e^-$ photoproduction.

\section*{Acknowledgments}
We are grateful to A.I. Milstein for the interest to the work and fruitful discussions.

\bibliography{ref}
\end{document}